\documentclass[10pt,prd,nofootinbib,english,twocolumn]{revtex4}
\usepackage{amsmath}
\usepackage{amsfonts,color}
\usepackage{xcolor}
\usepackage{amssymb,float}
\usepackage[utf8]{inputenc}
\usepackage{color}
\usepackage[unicode=true,pdfusetitle,
 bookmarks=true,bookmarksnumbered=true,bookmarksopen=true,bookmarksopenlevel=3,
 breaklinks=false,pdfborder={0 0 1},backref=false,colorlinks=true]
 {hyperref}
\usepackage{enumitem}
\usepackage{graphicx}
\usepackage{appendix}
\usepackage{mathrsfs}
\usepackage{hyperref}
\usepackage{soul} 
\usepackage[normalem]{ulem}
\usepackage{orcidlink}

\definecolor{v}{rgb}{0.6, 0.2, 0.8} 

\begin{document}

\title{Linear perturbations spectra for dynamical dark energy inspired by modified gravity}

\author{Celia Escamilla-Rivera}
\email{celia.escamilla@nucleares.unam.mx}
\affiliation{Instituto de Ciencias Nucleares, Universidad Nacional Aut\'onoma de M\'exico, Circuito Exterior C.U., A.P. 70-543, M\'exico D.F. 04510, M\'exico.}

\author{A. Hern\'andez-Almada}
\email{ahalmada@uaq.mx}
\affiliation{Facultad de Ingenier\'ia, Universidad Aut\'onoma de Quer\'etaro, Centro Universitario Cerro de las Campanas, 76010, Santiago de Quer\'etaro, M\'exico}

\author{Miguel A. Garc\'ia-Aspeitia}
\email{aspeitia@fisica.uaz.edu.mx}
\affiliation{Unidad Acad\'emica de F\'isica, Universidad Aut\'onoma de Zacatecas, Calzada Solidaridad esquina con Paseo a la Bufa S/N C.P. 98060, Zacatecas, M\'exico.}
\affiliation{Consejo Nacional de Ciencia y Tecnolog\'ia, \\ Av. Insurgentes Sur 1582. Colonia Cr\'edito Constructor, Del. Benito Ju\'arez C.P. 03940, Ciudad de M\'exico, M\'exico.}

\author{V. Motta}
\email{veronica.motta@uv.cl}
\affiliation{Instituto de F\'isica y Astronom\'ia, Facultad de Ciencias, Universidad de Valpara\'iso, Avda. Gran Breta\~na 1111, Valpara\'iso, Chile.}


\begin{abstract}
In this paper, we study a particular modified gravity Equation of State, the so-called Jaime-Jaber-Escamilla, that emerges from the first gravity modified action principle and can reproduce three cosmological viable $f(R)$ theories: the Starobinsky, Hu-Sawicki, and Exponential models. 
This EoS is a suitable candidate to reproduce the dynamical dark energy behaviour already reconstructed by the current data sets. Based on the joint statistical analysis, we found that our results are still in good agreement (within $1\sigma$) with the $\Lambda$CDM, while at perturbative level
we notice that the matter power spectrum normalisation factor $\sigma_8$ shows an agreement with SDSS and SNeIa+IRAS at 1-$\sigma$ for the Starobinsky model and with SDSS-vec for the Hu \& Sawicki and Exponential models. Furthermore, we found that for the $H_0$ values, Starobinsky and Hu \& Sawicki show  
the least tension in comparison with PL18 TT. All these aspects cannot be observed \textit{directly} from other alternatives theories, were a equation of state is difficult to compute analytically.
\end{abstract}

\maketitle
\section{Introduction} \label{Int}

One of the major challenges that precision cosmology face is a full explanation of the late cosmic acceleration. 
The standard concordance or Lambda Cold Dark Matter model ($\Lambda$CDM) offer an agreeable explanation where the observed accelerating expansion is related to the repulsive gravitational force of a Cosmological Constant $\Lambda$, with constant energy density $\rho$, and negative pressure $p$, having an equation of state (EoS) $\rho = -p$. However, even with this simple formulation, this concordance model suffers from severe theoretical inconveniences like the fine-tuning and coincidence problems \cite{Zeldovich,Weinberg}. To relax these inherent problems, several alternative proposals have been presented in the literature over the last years\footnote{ It is worth to mention that the quantum vacuum fluctuations (QVF) associated to the cosmological constant (the so-called fine-tuning problem) are still present in extensions to GR. However, it is assumed that these fluctuations are not the cause of the Universe acceleration as happens for the standard cosmology. This assumption alleviates the fine-tuning problem in the sense that it is easier to use a process to cancel all the contributions than to find a mechanism to obtain a small value of the energy density \cite{Weinberg}. A modern viewpoint of the problem of QVF and its relation with the cosmological constant can be seen in \cite{Juarez-Aubry:2019jbw}.} (see e.g. \cite{Escamilla-Rivera:2019ulu, Escamilla-Rivera:2019hqt, Hernandez-Almada:2018osh,Garcia-Aspeitia:2018fvw,Garcia-Aspeitia:2019yod,Hernandez-Almada:2020uyr,Garcia-Aspeitia:2019yni,Copeland:2006wr,amendola-book}) they usually propose modifications to General Relativity (GR) by considering an arbitrary function of the Ricci scalar $R$ as $f(R)$, with $f$ being an analytical function of $R$. Scenarios with a dynamical dark energy have been also considered. Both approaches got important attention because they provide explanations for several theoretical shortcomings \cite{DeFelice:2010aj,Jaime:2012nj}, but none of them can completely depict the evolution of the Universe. 

As a plausible solution, \cite{Jaime:2018ftn} showed that it is possible to construct a  $f(R)$ model using the chameleon mechanism that generates a late-time accelerating expansion of the Universe and is consistent with the Solar System constraints. The conditions for such viable $f(R)$ models include (i) the positivity of the effective gravitational coupling; (ii) the stability of cosmological perturbations; (iii) the stability of the late-time de-Sitter point; (iv) the asymptotic behaviour towards $\Lambda$CDM at the high curvature regime; (v) the Solar System constraints; and (vi) the constraint from the violation of the Equivalence Principle. Some of these conditions have been already pointed out in \cite{Jaime:2013zwa}.

At the moment, 
modified gravity can be comparable with the $\Lambda$CDM expansion history at an observationally equivalent level, while for the growth of structure there can be significant deviation that provide an important probe for MG.
Nevertheless, from a linear perturbative point of view, it should be possible to differentiate between them or at least find features that can deviate from $\Lambda$CDM, without 
considering a modification of the perturbed equations but instead  using the corresponding EoS for each $f(R)$ model.

The paper is organized as follows: In Sec. \ref{Theo} we give details about the $f(R)$ models and its relation with the Jaime-Jaber-Escamilla (JJE) equation of state. In Sec. \ref{Obsdat} the samples used to constraint the JJE parameterization are presented. Sec. \ref{Oconstr} is devoted to present the observational constraints at background and perturbative level. Finally in Sec. \ref{Con} we present the conclusions of the JJE parameterization, remarking the competitiveness with other parameterizations. Hereafter, we will use natural units in which $\hbar=c=k_B=1$, unless otherwise mentioned.

\section{Generic EoS for modified gravity} \label{Theo}

In these theories, we usually start with an action of the form
\begin{equation}
\label{f(R)}
S[g_{ab},{\mbox{\boldmath{$\psi$}}}] = \frac{1}{2\kappa}
\!\! \int \!\! d^4 x  f(R) \sqrt{-g} \: 
+ S_{\rm matt}[g_{ab}, {\mbox{\boldmath{$\psi$}}}] \; ,
\end{equation}
where $\kappa \equiv 8\pi G$, $S_{\rm matt}[g_{ab}, {\mbox{\boldmath{$\psi$}}}]$ is the standard 
matter action, {\mbox{\boldmath{$\psi$}}} represents the matter fields and  $f(R)$ is an arbitrary smooth function of the Ricci scalar $R$ and $g_{ab}$ is the metric tensor. Performing the variation of (\ref{f(R)}) we can obtain the field equations
\begin{eqnarray}
\label{fieldeq3}
G_{\mu\nu} &=& \frac{1}{f_R}\Bigl{[} f_{RR} \nabla_\mu \nabla_\nu R +
 f_{RRR} (\nabla_\mu R)(\nabla_\nu R) 
 \nonumber\\&&
-\frac{g_{\mu\nu}}{6}\Big{(} Rf_R+ f + 2\kappa T \Big{)} 
+ \kappa T_{\mu\nu} \Bigl{]}, \; 
\end{eqnarray}
where $G_{\mu\nu}= R_{\mu\nu}-g_{\mu\nu}R/2$ is the Einstein tensor, $T=T^{\alpha}_{\alpha}$ is the energy-momentum scalar and where $f_R\equiv\partial_Rf$, the other subscripts of R indicate higher orders in derivatives.

We are going to consider an homogeneous and isotropic Universe described by a flat ($k=0$) Friedman-Lema\^itre-Robertson-Walker (FLRW) metric
$ds^2 = - dt^2  + a^2(t)\!\left[ dr^2+ r^2 d\Omega^2\right]$, where $d\Omega^2=d\theta^2 + \sin^2\theta d\varphi^2$ is the solid angle. The energy-momentum tensor will be composed by baryons, dark matter (DM) and radiation, given by the equation $T_{\mu\nu}=Pg_{\mu\nu}+(\rho+P)u_{\mu}u_{\nu}$, where $P$, $\rho$ and $u_{\mu}$ are the pressure, the energy density and the quadri-velocity, respectively. Therefore, our set of evolution equations are given by
\begin{eqnarray}
\label{traceRt}
& \ddot R +3H \dot R = &-  \frac{1}{3 f_{RR}}\left[ 3f_{RRR} \dot R^2 + 2f- f_R R + \kappa T \right], \,\,\,\,\,\,\, \\
& H^2 = -\frac{1}{f_{RR}} & \left[f_{RR}H\dot{R}-\frac{1}{6}(Rf_{R}-f) \right]-\frac{\kappa T^{t}_{t}}{3f_{R}}, \label{eq:Hgen}\\
& \dot{H}+ H^2= &  -\frac{1}{f_{R}} \left[ f_{RR}H\dot{R} + \frac{f}{6}+\frac{\kappa T^{t}_{t}}{3} \right]  \,\,\,, \label{Hdotgen}
\end{eqnarray}
where $T^t_t$ refers to the temporal component of the energy-momentum tensor and $H\equiv\dot{a}/a$, while dots indicate derivatives with respect to the temporal coordinate. The previous equations are calculated in the traditional way, computing the Einstein tensor in the left side and calculating the covariant derivatives in the right side, leaving the functions $R$, and $f$ implicit, as well as the components of the energy-momentum tensor. 

The EoS for the dark energy fluid in $f(R)$ is given by $w_{\text{fld}} = (3H^2-3\kappa P-R)/[3(3H^2-\kappa\rho]$. The parameter $w_{\text{fld}}$ is the EoS for the geometric dark energy in $f(R)$ \cite{Jaime:2010kn} with 
a Ricci scalar given by $R = 6(\dot{H}+2H)$. The pressure and density are related to the matter and radiation content. As it was shown in \cite{Jaime:2018ftn}, there are three possible cosmological models: Starobinsky (Sta) model \cite{Starobinsky:2007},  Hu $\&$ Sawicki (HS) model \cite{Hu:2007} and the Exponential (Exp) model \cite{Linder:2009,Jaime:2017gtr,Odintsov:2017qif}. To obtain a generic reconstruction for the EoS adapted to $f(R)$ cosmological viable models  we perform numerical integration
of  these field equations to obtain the Jaime-Jaber-Escamilla EoS for dark energy surveys \cite{Jaime:2018ftn}:
\begin{equation}\label{eq:JJE_original}
w(z) = -1 + \frac{w_0}{1+w_1z^{w_2}}\cos(w_3+z),
\end{equation}
which can reproduce the following cosmological viable models at background level:
\begin{equation}
\begin{cases}
 f(R)_{\text{Sta}}= R+\lambda R_{Sta}\left[ \left( 1+\frac{R^2}{R^2_{Sta}}\right)^{-q}-1\right], \\ \\
f(R)_{\text{HS}}= R- R_{\rm HS}\frac{c_{1}\left(\frac{R}{R_{\rm HS}}\right)^n}{c_{2}\left(\frac{R}{R_{\rm HS}}\right)^n+1} \; ,\\ \\
 f(R)_{\text{Exp}}=R+\beta R_* (1-e^{-R/R_*}),
   \end{cases}
\end{equation}
where $R_{Sta}$, $R_{HS}$ and $R_*$ are appropriate constants for each model, $w_i$ ($i=0,1,2,3$), $q$, $c_1$, $c_2$, $n$, and  $\beta$ are free parameters, $z$ is the standard redshift given by $z=a_0/a-1$ and $a_0$ is normalized to one. Notice that (\ref{eq:JJE_original}) has a current value given by $w(z=0) = w_0\cos(w_3)-1$, which can recovers $w=-1$ at $z\gg 1$, and allows dynamical oscillations in the 
redshift range of current and future surveys \cite{delaTorre:2016rxm,Schlegel:2019eqc,Bowman:2018yin,LIGOScientific:2020stg,Corman:2020pyr}.  According to the numerical solution of the field equations, the evolution of each of them can be recovered within a $0.5\%$ for the first two, and within a $0.8\%$ precision for the latter. 
These percentages are quite reasonable values since the estimated accuracy of current and future experiments  provide a statistical significance of $1\%$ below $z=1$ for the JJE parameterization.

The key advantage of using the Lagrangian formalism from $f(R)$, as opposed to an ad-hoc approach at the level of the field and fluid equations, is self-consistency. This approach leads to dynamically evolving $w$ and $c^{2}_{s}$ that are derived from an action. This procedure also avoids unforeseen instabilities since usual pathologies like ghost and strong coupling problems can be immediately identified from the action. When discussing effective models of dark energy, the focus is $w$ and how it affects observations such as the CMB and the matter power spectrum. However, another parameter of interest is the dark energy speed of sound, $c_s$, and its observational signatures and constraints using various cosmological probes.
Although this parameter remains unconstrained by observations, future cosmological experiments such as Euclid or SKA could constrain it. Therefore, it is important to fully explore the possibility of a varying $c_s$ using a well formulated theory. In our case we consider $c^2_{s}=1$.

In this paper, we explore a generic form of JJE EoS with fixed values that is able to pass the Solar System tests \cite{Capozziello:2007eu, Jaime:2018ftn}, i.e. these tests alone can place weak constraints on these $f(R)$-like models, since the additional scalar degree of freedom is locked to the high-curvature general-relativistic prediction across more than 25 orders of magnitude in density, throughout the solar corona. This requires a sufficiently extent galactic halo to maintain the galaxy at high curvature in the presence of the low-curvature cosmological background.
This goal can be obtained by rewriting (\ref{eq:JJE_original}) as:
\begin{equation}\label{eq:JJE}
w(z)=-1+\frac{w_a \cos (\alpha \nu(z) )}{w_b z^p+1},
\end{equation}
where $\nu (z) = 2\pi/(\sqrt{\eta}z+1)^{1/2}$ and $w_a$, $w_b$, $\alpha$, $\eta$, and $p$ are constants\footnote{The relationship between parameters in Eqs.(\ref{eq:JJE_original})-(\ref{eq:JJE}) is: $w_0 = w_a$, which is solely an amplitude, $w_1 = w_b$ and $p=w_2$, re-labeled to differentiate them from the standard ones in JJE parameterisation. Furthermore, the argument in the cosine $(w_3 + z)$, was rewritten as $\alpha \nu(z)$ given that cases with periodic behaviours are consistent with Solar system tests when $\cos{\nu (z) = 2\pi/(\sqrt{6}z+1)^{1/2}}$.}

To be consistent with the solar observations, we have set $w_b=0.03$, $\alpha=1$, $\eta=6$ according to the values for each $f(R)$ model given in \cite{Hu:2007nk,Linder:2009jz,Odintsov:2017qif}. 
 We consider $p=11$ for the background analysis since it can reproduce the Exponential model and, according to the current astrophysical data, it converges faster in comparison to the other two models. For the linear perturbations, we consider three values $p=\{4,5,11\}$, which are in agreement with the cosmological viable cases \cite{Jaime:2018ftn}. Notice that, according to the original JJE parametrization (\ref{eq:JJE}),  these values for $p$ correspond
to the numerical solutions that reproduce the three possible cosmological models derived from modified gravity. 

\section{Observational data} \label{Obsdat}

To perform the statistical analyses of (\ref{eq:JJE})  and understand current constraints, we need to focus on specific data sets and likelihood functions. For this purpose we consider the following first four samples for the background analysis and, all of them and their combination with Planck 2018 for the perturbation.

\begin{itemize}

\item Pantheon SNeIa binned compilation: we consider the 40 bins compressed from 1048 SNeIa in the redshift range $z \in [0.01, 2.3]$ \cite{Scolnic:2017caz}. 

\item BAO measurements: we consider the sample of 6 correlated data points, with their associated covariance matrix, collected in \cite{Giostri:2012} and measured by \cite{Percival:2010,Blake:2011,Beutler:2011hx}.

\item Hubble parameter measurements (H(z)): we consider a sample of 31 model-independent measurements which use the differential age method proposed by \cite{Jimenez:2001gg}.

\item Strong Lensing (SLS): we consider a catalog with 205 systems in a redshift range $0.0625 < z_{l}< 0.958$ for the lens galaxy and $0.196<z_s < 3.595$ for the source \cite{Amante:2019xao} (see also \cite{Biesiada:2006zf,Biesiada:2010,Cao:2012,Cao:2015qja}).

\item Planck Legacy 2018 (PL18):  we adopt the low-$l$ and high-$l$ likelihoods from  \cite{Planck_overview:2018, Planck_like:2018}, from the temperature power spectra (TT).

\end{itemize}

\section{Observational Constraints} \label{Oconstr}

\begin{itemize}
\item Background level: The proposed EoS  set a formulation that can be systematically implemented in several surveys to test for alternative theories of gravity. Therefore, we can  find the best-fit values for the free cosmological parameters using the samples described previously. These values will be used for the linear perturbations. We perform MCMC and Bayesian analyses 
and report the best-fit values for the generic EoS (\ref{eq:JJE}) for the case $p=11,5,4$ and LCDM in Table \ref{tab:bf_background}. Figure 1 displays the 2D phase-space of the free parameters $(h,\Omega_{m0},w_a)$ at $68\% (1\sigma)$, $95\%(2\sigma)$, and $99.7\%(3\sigma)$ confidence level (C.L.),  as well as the 1D posterior distributions using the individual and joint samples. According to the $\chi^2$-value, we find a good statistical agreement with the data density. 
In our generic EoS (\ref{eq:JJE}) we  set $p$ as free parameter  to find flat posterior distributions, allowing  a negligible difference between $f(R)$ models. Figure 2 shows
the reconstruction of $w(z)$  for each observable up to $3\sigma$. 

\begin{table*}
\centering
\begin{tabular}{|c|cccc|}
\hline
Sample      &  $\chi^2$-value     &  $h$ & $\Omega_{m0}$ & $w_a$           \\
\hline
\multicolumn{5}{|c|}{$f(R)$ model: p=11} \\
\hline
H(z) &	   $13.9$      & $0.740^{+0.017}_{-0.016}$ & $0.244^{+0.038}_{-0.035}$ & $-0.502^{+0.301}_{-0.298}$  \\  [0.7ex]
Pantheon SNeIa &   $48.4$     & $0.740^{+0.017}_{-0.017}$ & $0.305^{+0.037}_{-0.036}$ & $-0.043^{+0.163}_{-0.189}$   \\  [0.7ex]
SLS   &  $590.9$    & $0.740^{+0.017}_{-0.017}$ & $0.092^{+0.045}_{-0.039}$ & $-1.010^{+0.293}_{-0.266}$   \\  [0.7ex]
BAO & $2.8$ & $0.740^{+0.017}_{-0.017}$ & $0.296^{+0.08}_{-0.07}$ & $-0.013^{+0.932}_{-0.848}$  \\  [0.7ex]
Joint    & $650.2$ & $0.724^{+0.013}_{-0.013}$ & $0.280^{+0.016}_{-0.015}$ & $-0.074^{+0.102}_{-0.105}$  \\  [0.7ex]
\hline
\hline
\multicolumn{5}{|c|}{$f(R)$ model: p=5} \\
\hline
Joint   & $650.2$ & $0.724^{+0.013}_{-0.013}$ & $0.280^{+0.016}_{-0.016}$ & $-0.074^{+0.102}_{-0.105}$  \\  [0.7ex]
\hline
\hline
\multicolumn{5}{|c|}{$f(R)$ model: p=4} \\
\hline
Joint   & $650.2$ & $0.724^{+0.013}_{-0.013}$ & $0.280^{+0.016}_{-0.016}$ & $-0.076^{+0.103}_{-0.104}$  \\  [0.7ex]
\hline
\hline
\multicolumn{5}{|c|}{$\Lambda$CDM} \\
\hline
Joint   & $650.8$ & $0.723^{+0.013}_{-0.013}$ & $0.278^{+0.016}_{-0.015}$ & --  \\  [0.7ex]
\hline
\end{tabular}
\caption{Background best-fit values for Eq. (\ref{eq:JJE}) with $w_b = 0.03$, $\eta= 6$, $\alpha=1$ and $p=11,\,5,\,4$, where the joint sample indicates H(z)+Pantheon+SLS+BAO. The normalised Hubble value $h$ is also reported.}
\label{tab:bf_background}
\end{table*}

\begin{table*}
\centering
\begin{tabular}{|c|cccc|}
\hline
 & &\multicolumn{3}{c|}{{$f(R)$ viable models}}
\\ \hline
Parameters & $\Lambda$CDM & Starobinsky ($p=5$)  &  Hu \& Sawicki ($p=4$) & Exponential ($p=11$)
\\ \hline
$H_0$ &$67.69^{+1.14}_{-1.24}$ & $ 68.10^{+1.18}_{-1.10} $ & $68.22^{+1.45}_{-2.11}$ & $71.22^{+1.05}_{-2.00}$ 
\\ \hline
100$\Omega_b h^2$ & $ 2.23 \pm 0.03 $ & $2.23 \pm 0.03 $ & $2.31^{+0.11}_{-1.02}$ & $2.43^{+1.22}_{-1.12}$ 
\\ \hline
$\Omega_c h^2$ & $0.119 \pm 0.002$ & $0.118 \pm 0.022$ & $0.118 \pm 0.003$  & $0.119 \pm 0.013$ 
\\ \hline
$\tau$ & $0.059^{+0.041}_{-0.101}$ &  $ 0.049\pm 0.032 $ &$0.049^{+0.132}_{-0.111}$ &$ 0.071 \pm 0.038 $ 
\\ \hline
$\mathrm{ln}(10^{10} A_s)$ &  $3.071^{+0.054}_{-0.052}$ & $ 3.047^{+0.057}_{-0.059} $ & $3.026 \pm 0.061$  & $3.026 \pm 0.061$  
\\ \hline
$n_s$ &  $ 0.970^{+0.008}_{-0.007} $ & $0.969 \pm 0.007$ & $0.970 \pm 0.008$ & $0.975\pm{0.118}$ 
\\ \hline
$\sigma_8$ &  $0.811^{+0.025}_{-0.026}$ & $ 0.811^{+0.047}_{-0.075} $ & $0.942\pm 0.041$  & $0.945^{+0.140}_{-0.045}$  \\ \hline
 $\chi^2_{\text{min}}$  &  1276.567 &  1436.434  & 1156.243  & 1341.232  \\
\hline 
\end{tabular}
\caption{Results from the linear perturbation analysis ($95\%$ C.L.) for the $\Lambda$CDM and the three cosmological viable $f(R)$ gravity models.}
\label{tab:perturbations_data}
\end{table*}
 
\begin{figure}
\centering
\includegraphics[width=0.5\textwidth,origin=c,angle=0]{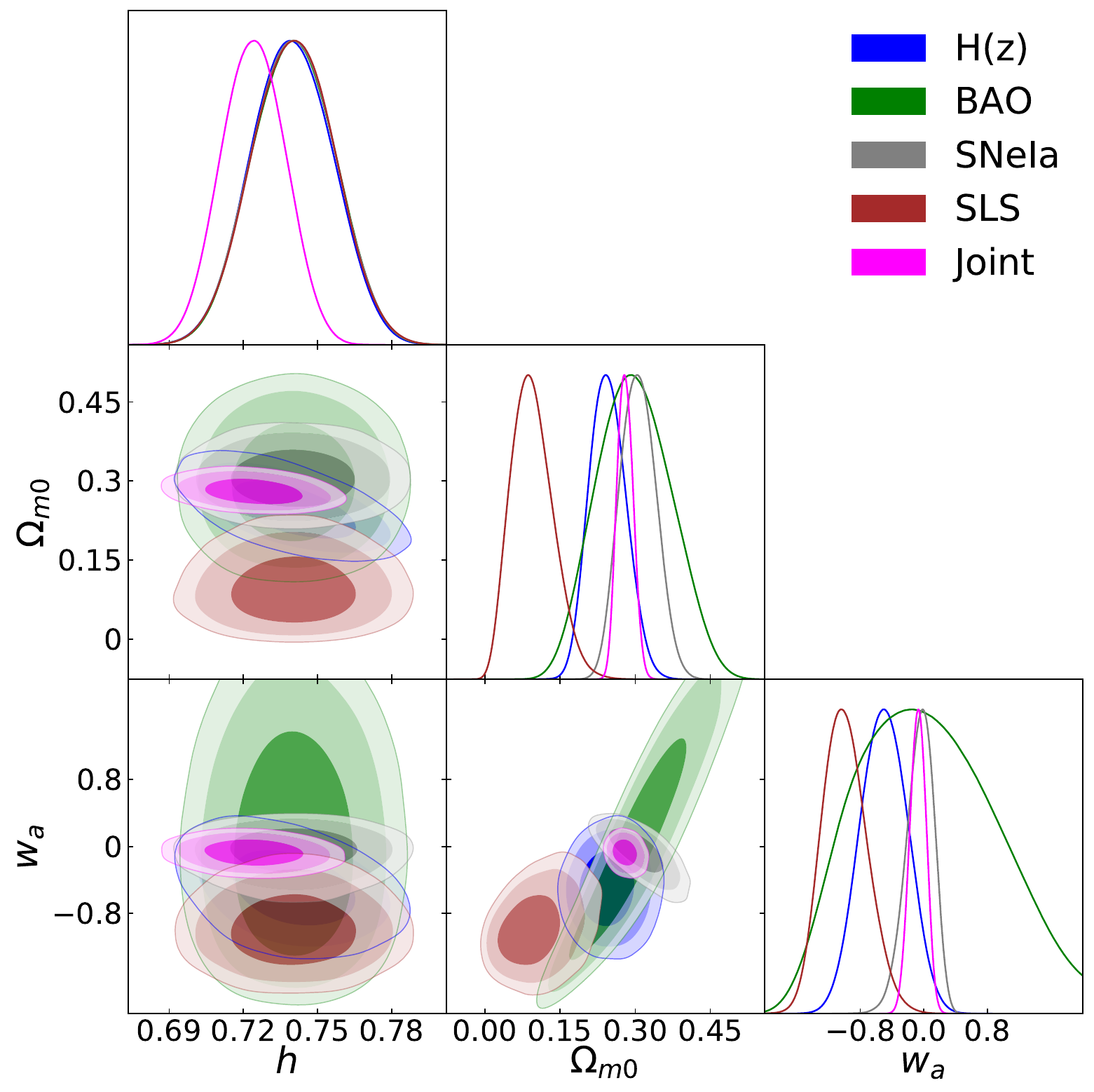}
\caption{The regions show the 68\%, 95\% and 99.7\% confidence level (C.L.) contours inferred from (\ref{eq:JJE}) with $w_b = 0.03$, $\eta= 6$, $\alpha=1$ and $p=11$. For this latter value we perform a MCMC calculations with $p=4$, which shows a posterior evidence less than $0.001\%$ according to \cite{Jaime:2018ftn} in comparison to $p=11$. We choose this latter case to approximate the $H_0$ value from late Universe measurements.}
\label{fig:contour}
\end{figure}

\begin{figure*}
\centering
     \includegraphics[width=0.4\textwidth]{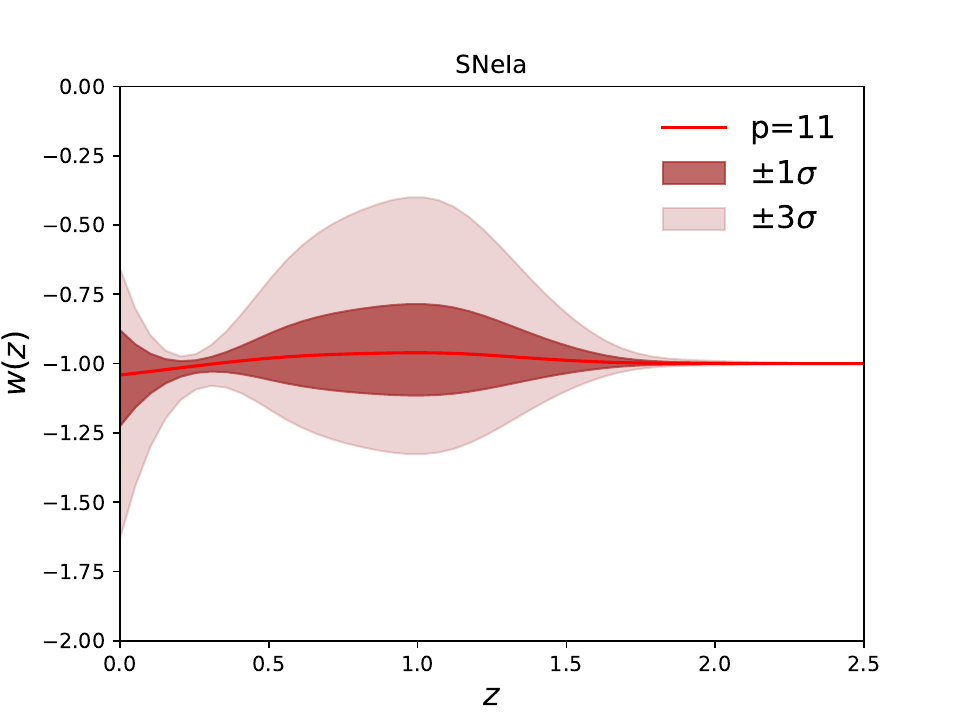}
     \includegraphics[width=0.4\textwidth]{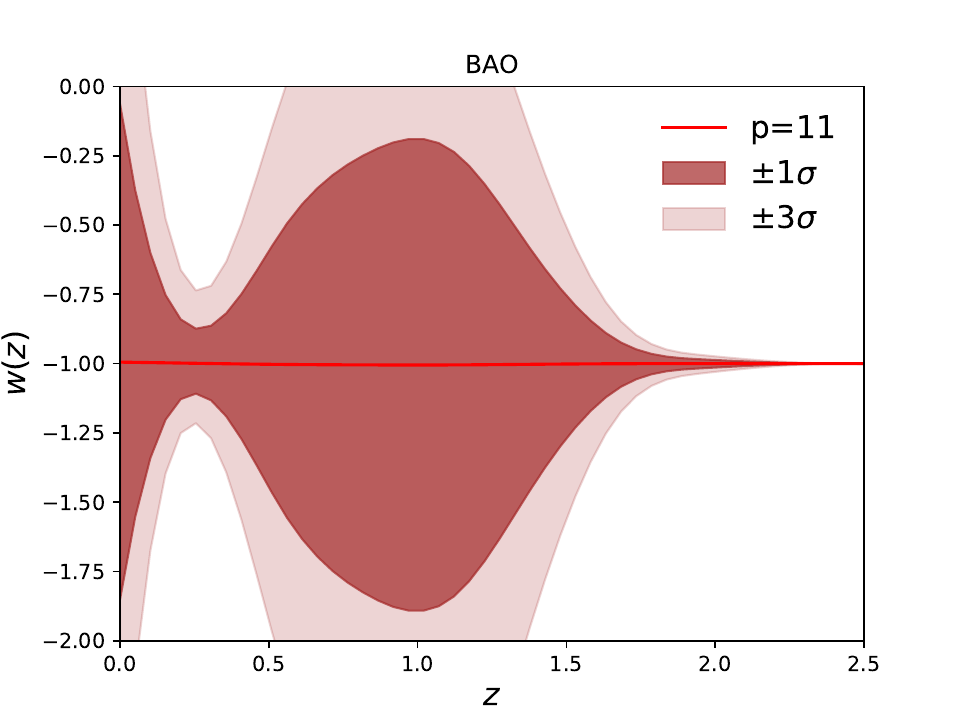} \\
     \includegraphics[width=0.4\textwidth]{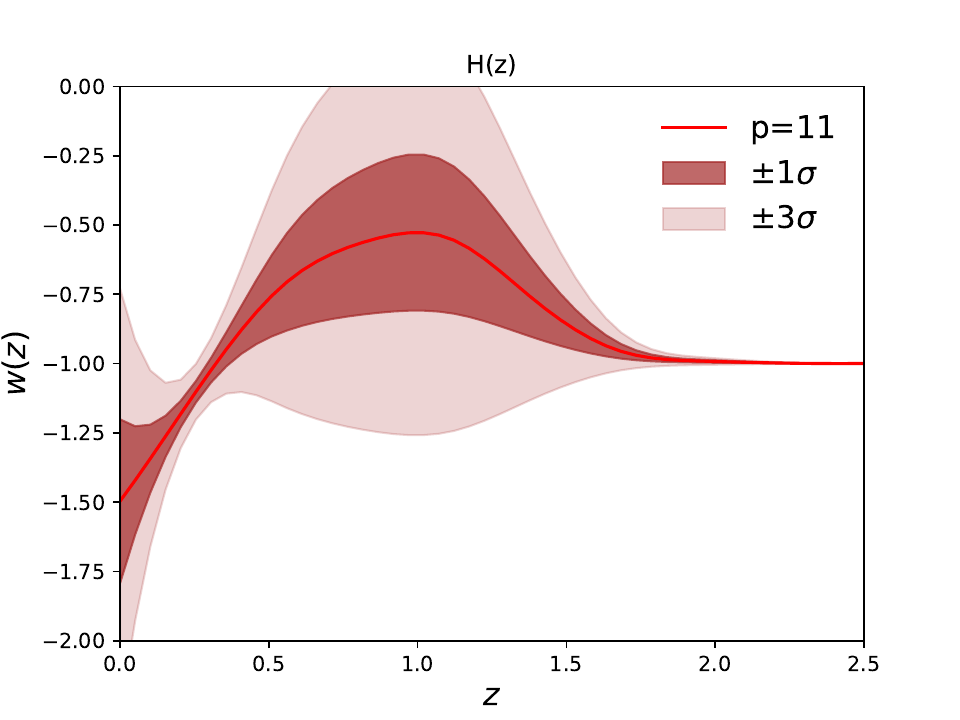}
     \includegraphics[width=0.4\textwidth]{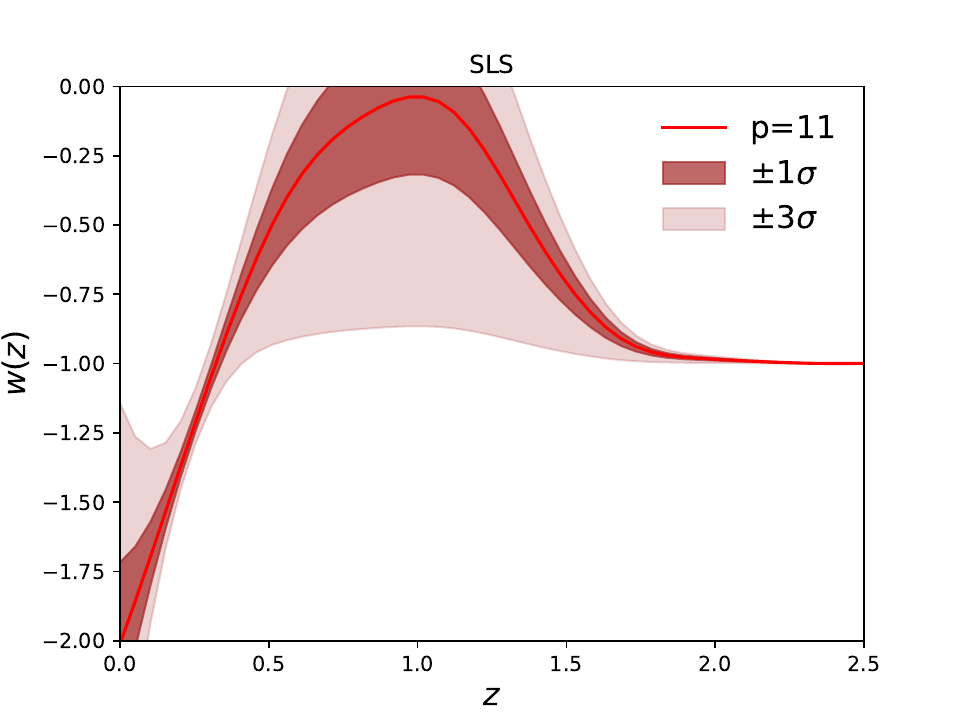} \\
     \includegraphics[width=0.4\textwidth]{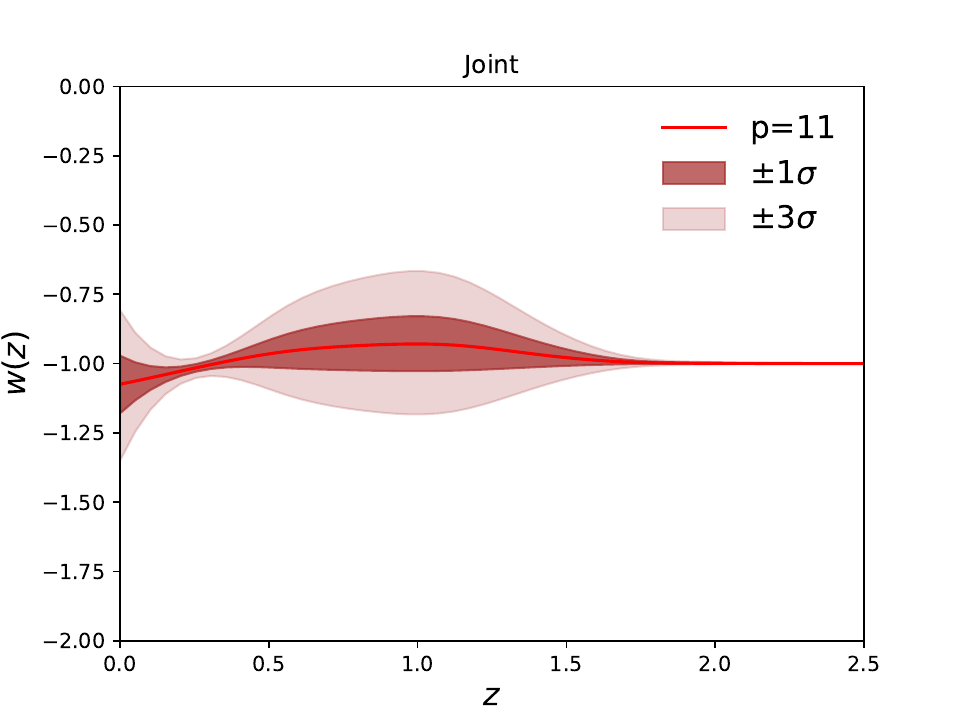}
    \caption{Dynamical EoS $w$ 
    given by (\ref{eq:JJE}) with $w_b = 0.03$, $\eta= 6$, $\alpha=1$ and $p=11$. Darker (lighter) bands represent the uncertainty at $1\sigma$($3\sigma$) C.L.
We illustrated the calculations for each observational samples: 
(\textit{Top} to \textit{Bottom}, \textit{Left} to \textit{Right}):
Pantheon SNeIa, BAO, H(z), 
SLS and the full (joint) data.}
    \label{fig:EoS}
\end{figure*}

\begin{figure}
\centering
\includegraphics[width=0.5\textwidth,origin=c,angle=0]{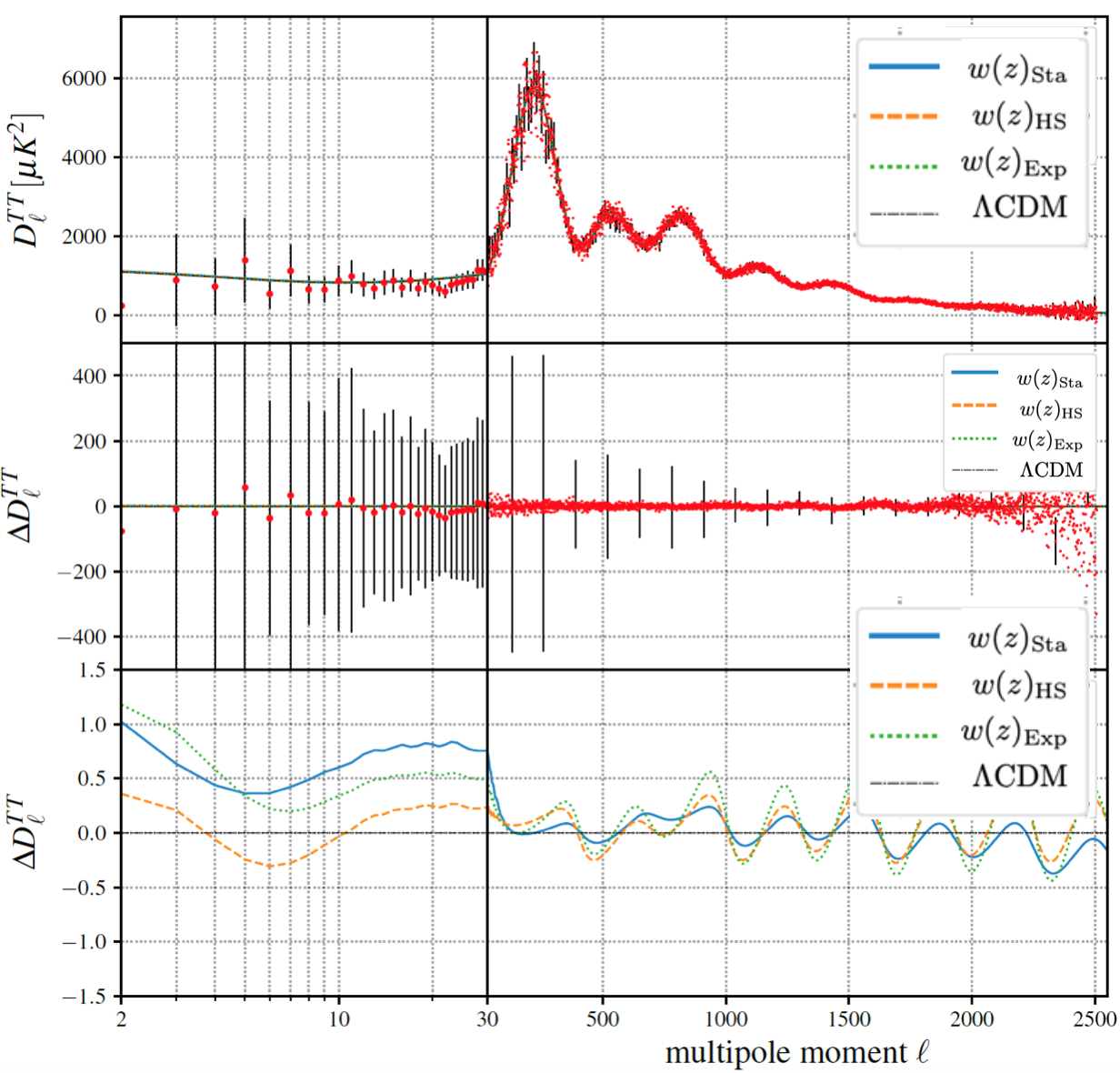}
\caption{The CMB $C^{TT}_{l}$ power spectra versus multipole moment $l$ using the best fits reported in Table II for the joint sample CMB+BAO+SLS+H(z)+Pantheon SNeIa. The $f(R)$ models are represented by: Starobinksy (solid blue), Hu \& Sawicky (dashed orange) and Exponential (dotted green). The red dots with black error bars indicate PL18 TT. \textit{Top:} $D^{TT}_{l}$ fit. \textit{Middle:} Percentage difference with respect to $\Lambda$CDM (dotted-dashed black). \textit{Bottom:} Percentage difference according to deviations from $\Lambda$CDM.
} 
\label{fig:pert_general}
\end{figure}

\item Linear perturbation spectra for $f(R)$: This is a novel attempt to present the treatment of the linear evolution of cosmological viable $f(R)$ models from the inflation era until late times. To achieve this goal, we 
have implemented the JJE EoS parametrization in CLASS
\footnote{\url{https://lesgourg.github.io/class_public/class.html}}. 
We start with a specific cosmological model (in our case, the three $f(R)$ scenarios with a generic EoS). We begin by numerically solving the Friedmann equations for the cosmic background with $w_{i}$ mimicking a dark fluid and, at a subsequent time, solving the thermodynamic evolution of the system. Afterward, we obtain the primordial power spectrum $P=A_s (k/k_{*})^{n_s -1}$. In addition, considering that modified gravity can produce degeneracy effects in the total mass of the neutrinos, we have set the sum of the neutrino masses as 0.06 eV. The advantage of using a generic EoS (eq. \ref{eq:JJE}) 
is that  $f(R)$ gravity\footnote{Codes like HiCLASS offer a possibility to solve high order theories, e.g. Hordenski theories.}  can be easily implemented at a background level and used as any dark energy-like proposal at standard EoS level. 
To break the degeneracy,  we consider all the samplers used at background level, i.e. we combine BAO+SLS+H(z)+Pantheon  SNeIa together with the CMB data from the final release of the Planck collaboration (2018). The comparison of the results
are presented in Table \ref{tab:perturbations_data} and Figure 3.
All the models present at log power of the primordial curvature perturbations and a scalar index $n_s$ with $k_0 =0.05 \text{Mpc}^{-1}$. The $\Delta D^{TT} = D^{TT}_{l~\text{model}} -D^{TT}_{l~\Lambda CDM}$ \cite{Escamilla-Rivera:2016aca}
shows a deviation from $\Lambda$CDM between [0.5\%-0.8\%] at low-\textit{l}, which is in agreement with the interval precision to reproduce $f(R)$-like evolution over the numerical solutions of the field equations \eqref{traceRt}-\eqref{Hdotgen}. 
\end{itemize}

\section{Conclusions} \label{Con}

We devoted our work to the analysis of a dynamical dark energy model that not only mimic $f(R)$ theories at the background level but also exhibits perturbations behaviour different than other $f(R)$ theories.
The background fitting analysis was developed from a generic EoS that can reproduce these models.
In this approach, an optimal value for $p$ is extracted from the cosmological viable  values (4,5,11) in (\ref{eq:JJE}). The case $p=11$  reproduces our three $f(R)$ models and pass the Solar system test, therefore reducing the statistical phase space. Furthermore, $w_a$ (which mimics $w_0$ from the original JJE EoS proposal) acts as an amplitude in the second term of Eq.(\ref{eq:JJE}). The parameters in this equation are constrained by observational data, which in turn affect the posterior probabilities as we can see from Fig. 1. A deviation from $\Lambda$CDM occurs when $w_a \neq0$,allowing us to measure the deviation between LCDM and the modified gravity scenarios.

 Our analysis shows the possibility of constraining an effective theory of dynamical dark energy that can mimic $f(R)$ theories at the background level. On the other hand, at linear perturbation level, they show differences with other $f(R)$ modified theories of gravity.
Based on the joint statistical analysis, we found that (\ref{eq:JJE}) is still in good agreement (within $1\sigma$) with the concordance model.
Afterwards, we study the linear perturbations by considering our proposal (\ref{eq:JJE}) and the three $f(R)$ scenarios. We notice that the matter power spectrum normalisation factor $\sigma_8$ shows an agreement with SDSS and SNeIa+IRAS at 1-$\sigma$ for the Starobinsky model and with SDSS-vec for the Hu \& Sawicki and Exponential models according to \cite{Kazantzidis:2018rnb}. We also find that, for the $H_0$ values, Starobinsky and Hu \& Sawicki show  
the least tension in comparison with PL18 TT, while the exponential model indicates at 2\%-tension with other cosmological measurements such as, for example, strong lensing time delays (H0LiCOW) \cite{Verde:2019ivm}. There is also evidence of $\gtrsim 2\sigma$ tension between the constraints from Planck on the matter density $\Omega_m$ and the amplitude $\sigma_8$ of matter fluctuations in linear theory and those from local measurements  \cite{Battye:2014qga,McCarthy:2017csu,DiValentino:2020vvd}. 
Using Planck we can derive $S_8 = \sigma_8 (\Omega_m/0.3)^{0.5} = 0.832\pm0.013$ whereas local measurements find the smaller values as: 
$S_8^{\rm SZ}=\sigma_8(\Omega_m/0.27)^{0.3}=0.78\pm0.01$ from Sunyaev-Zeldovich cluster counts \citep{2014A&A...571A..20P}, 
$S_8=0.783^{+0.021}_{-0.025}$ from DES  \citep{2017arXiv170801530D}
and $S_8=0.745 \pm 0.039$ from KiDS-450 \cite{Hildebrandt:2016iqg} weak-lensing surveys. 
The CFHTLenS weak-lensing survey also finds support for disagreement with Planck CMB predictions \cite{Joudaki:2016mvz}. According to these values, our results reported in Table II seems to point out an agreement with Planck.

On their own, SNeIa are poor probes of the absolute distance scale of the  Universe (and hence $H_0$). In our analysis, where an anchor at high redshift is provided by time-delay distances of strong lensing, their main role is to extrapolate these absolute distance measurements back to redshift zero. JJE EoS allows us to constrain $H_0$ in a way that is rather insensitive to the assumed cosmological background model and independent of Cepheids and the CMB data.
Furthermore, one of the problems of $f(R)$ scenarios is the usual high value for $\sigma_8$ in the early Universe when considering RSD data. We will report this analysis in future works elsewhere. In fact, we have fixed all parameter values equal to those values of LCDM because we are interested  in finding a deviation of our EoS  from $w=-1$, in this sense, $w_a$ gives us such deviation.
Moreover, using the amplitude of matter fluctuations we notice that our results from  Hu \& Sawicki and Exponential can mimic a $\Lambda$CDM + $\gamma_{p}$ model \cite{Zhao:2017jma}. 
Furthermore, the equally impressive quantity of different viable candidates can also arise a confusion between models, at this point independent analyses are viable paths modeled as cosmographic solutions \cite{Capozziello:2008qc,Capozziello:2019cav,Capozziello:2015rda}.

Finally, from our general results we can conclude that our three $f(R)$ scenarios derived from (\ref{eq:JJE_original}) are compatible with $\Lambda$CDM at 2-$\sigma$ C.L. and at perturbative level show less tension with the samples described.

\begin{acknowledgements}
CE-R acknowledges the Royal Astronomical Society as FRAS 10147, supported by DGAPA-PAPIIT-UNAM Project IA100220 and would like to acknowledge networking support by the COST Action CA18108.
M.A.G.-A. acknowledges support from SNI-M\'exico, CONACyT research fellow, COZCyT and Instituto Avanzado de Cosmolog\'ia (IAC) collaborations. 
A.H.A. thanks to the support from Luis Aguilar, Alejandro de Le\'on, Carlos Flores, and Jair Garc\'ia of the Laboratorio Nacional de Visualizaci\'on Cient\'ifica Avanzada.
V.M. acknowledges the support of Centro de Astrof\'{\i}sica de Valpara\'{\i}so (CAV). M.A.G.-A and V.M. acknowledge the support of project CONICYT REDES (190147). We thank the anonymous referees for thoughtful remarks and suggestions.
\end{acknowledgements} 

\bibliography{librero0}

\begin{thebibliography}{57}
\expandafter\ifx\csname natexlab\endcsname\relax\def\natexlab#1{#1}\fi
\expandafter\ifx\csname bibnamefont\endcsname\relax
  \def\bibnamefont#1{#1}\fi
\expandafter\ifx\csname bibfnamefont\endcsname\relax
  \def\bibfnamefont#1{#1}\fi
\expandafter\ifx\csname citenamefont\endcsname\relax
  \def\citenamefont#1{#1}\fi
\expandafter\ifx\csname url\endcsname\relax
  \def\url#1{\texttt{#1}}\fi
\expandafter\ifx\csname urlprefix\endcsname\relax\def\urlprefix{URL }\fi
\providecommand{\bibinfo}[2]{#2}
\providecommand{\eprint}[2][]{\url{#2}}

\bibitem[{\citenamefont{Zel'dovich}(1968)}]{Zeldovich}
\bibinfo{author}{\bibfnamefont{Y.~B.} \bibnamefont{Zel'dovich}},
  \bibinfo{journal}{Soviet Physics Uspekhi} \textbf{\bibinfo{volume}{11}},
  \bibinfo{pages}{381} (\bibinfo{year}{1968}).

\bibitem[{\citenamefont{Weinberg}(1989)}]{Weinberg}
\bibinfo{author}{\bibfnamefont{S.}~\bibnamefont{Weinberg}},
  \bibinfo{journal}{Reviews of Modern Physics} \textbf{\bibinfo{volume}{61}}
  (\bibinfo{year}{1989}).

\bibitem[{\citenamefont{Ju\'arez-Aubry}(2019)}]{Juarez-Aubry:2019jbw}
\bibinfo{author}{\bibfnamefont{B.~A.} \bibnamefont{Ju\'arez-Aubry}},
  \bibinfo{journal}{Phys. Lett. B} \textbf{\bibinfo{volume}{797}},
  \bibinfo{pages}{134912} (\bibinfo{year}{2019}), \eprint{1903.03924}.

\bibitem[{\citenamefont{Escamilla-Rivera and
  Levi~Said}(2019)}]{Escamilla-Rivera:2019ulu}
\bibinfo{author}{\bibfnamefont{C.}~\bibnamefont{Escamilla-Rivera}}
  \bibnamefont{and} \bibinfo{author}{\bibfnamefont{J.}~\bibnamefont{Levi~Said}}
  (\bibinfo{year}{2019}), \eprint{1909.10328}.

\bibitem[{\citenamefont{Escamilla-Rivera
  et~al.}(2020)\citenamefont{Escamilla-Rivera, Quintero, and
  Capozziello}}]{Escamilla-Rivera:2019hqt}
\bibinfo{author}{\bibfnamefont{C.}~\bibnamefont{Escamilla-Rivera}},
  \bibinfo{author}{\bibfnamefont{M.~A.~C.} \bibnamefont{Quintero}},
  \bibnamefont{and}
  \bibinfo{author}{\bibfnamefont{S.}~\bibnamefont{Capozziello}},
  \bibinfo{journal}{JCAP} \textbf{\bibinfo{volume}{03}}, \bibinfo{pages}{008}
  (\bibinfo{year}{2020}), \eprint{1910.02788}.

\bibitem[{\citenamefont{Hernandez-Almada
  et~al.}(2019)\citenamefont{Hernandez-Almada, Magana, Garcia-Aspeitia, and
  Motta}}]{Hernandez-Almada:2018osh}
\bibinfo{author}{\bibfnamefont{A.}~\bibnamefont{Hernandez-Almada}},
  \bibinfo{author}{\bibfnamefont{J.}~\bibnamefont{Magana}},
  \bibinfo{author}{\bibfnamefont{M.~A.} \bibnamefont{Garcia-Aspeitia}},
  \bibnamefont{and} \bibinfo{author}{\bibfnamefont{V.}~\bibnamefont{Motta}},
  \bibinfo{journal}{Eur. Phys. J. C} \textbf{\bibinfo{volume}{79}},
  \bibinfo{pages}{12} (\bibinfo{year}{2019}), \eprint{1805.07895}.

\bibitem[{\citenamefont{Garcia-Aspeitia
  et~al.}(2018)\citenamefont{Garcia-Aspeitia, Hernandez-Almada, Magaña,
  Amante, Motta, and Martínez-Robles}}]{Garcia-Aspeitia:2018fvw}
\bibinfo{author}{\bibfnamefont{M.~A.} \bibnamefont{Garcia-Aspeitia}},
  \bibinfo{author}{\bibfnamefont{A.}~\bibnamefont{Hernandez-Almada}},
  \bibinfo{author}{\bibfnamefont{J.}~\bibnamefont{Magaña}},
  \bibinfo{author}{\bibfnamefont{M.~H.} \bibnamefont{Amante}},
  \bibinfo{author}{\bibfnamefont{V.}~\bibnamefont{Motta}}, \bibnamefont{and}
  \bibinfo{author}{\bibfnamefont{C.}~\bibnamefont{Martínez-Robles}},
  \bibinfo{journal}{Phys. Rev. D} \textbf{\bibinfo{volume}{97}},
  \bibinfo{pages}{101301} (\bibinfo{year}{2018}), \eprint{1804.05085}.

\bibitem[{\citenamefont{García-Aspeitia
  et~al.}(2019{\natexlab{a}})\citenamefont{García-Aspeitia, Hernández-Almada,
  Magaña, and Motta}}]{Garcia-Aspeitia:2019yod}
\bibinfo{author}{\bibfnamefont{M.~A.} \bibnamefont{García-Aspeitia}},
  \bibinfo{author}{\bibfnamefont{A.}~\bibnamefont{Hernández-Almada}},
  \bibinfo{author}{\bibfnamefont{J.}~\bibnamefont{Magaña}}, \bibnamefont{and}
  \bibinfo{author}{\bibfnamefont{V.}~\bibnamefont{Motta}}
  (\bibinfo{year}{2019}{\natexlab{a}}), \eprint{1912.07500}.

\bibitem[{\citenamefont{Hernández-Almada
  et~al.}(2020)\citenamefont{Hernández-Almada, Leon, Magaña,
  García-Aspeitia, and Motta}}]{Hernandez-Almada:2020uyr}
\bibinfo{author}{\bibfnamefont{A.}~\bibnamefont{Hernández-Almada}},
  \bibinfo{author}{\bibfnamefont{G.}~\bibnamefont{Leon}},
  \bibinfo{author}{\bibfnamefont{J.}~\bibnamefont{Magaña}},
  \bibinfo{author}{\bibfnamefont{M.~A.} \bibnamefont{García-Aspeitia}},
  \bibnamefont{and} \bibinfo{author}{\bibfnamefont{V.}~\bibnamefont{Motta}}
  (\bibinfo{year}{2020}), \eprint{2002.12881}.

\bibitem[{\citenamefont{García-Aspeitia
  et~al.}(2019{\natexlab{b}})\citenamefont{García-Aspeitia, Martínez-Robles,
  Hernández-Almada, Magaña, and Motta}}]{Garcia-Aspeitia:2019yni}
\bibinfo{author}{\bibfnamefont{M.~A.} \bibnamefont{García-Aspeitia}},
  \bibinfo{author}{\bibfnamefont{C.}~\bibnamefont{Martínez-Robles}},
  \bibinfo{author}{\bibfnamefont{A.}~\bibnamefont{Hernández-Almada}},
  \bibinfo{author}{\bibfnamefont{J.}~\bibnamefont{Magaña}}, \bibnamefont{and}
  \bibinfo{author}{\bibfnamefont{V.}~\bibnamefont{Motta}},
  \bibinfo{journal}{Phys. Rev. D} \textbf{\bibinfo{volume}{99}},
  \bibinfo{pages}{123525} (\bibinfo{year}{2019}{\natexlab{b}}),
  \eprint{1903.06344}.

\bibitem[{\citenamefont{Copeland et~al.}(2006)\citenamefont{Copeland, Sami, and
  Tsujikawa}}]{Copeland:2006wr}
\bibinfo{author}{\bibfnamefont{E.~J.} \bibnamefont{Copeland}},
  \bibinfo{author}{\bibfnamefont{M.}~\bibnamefont{Sami}}, \bibnamefont{and}
  \bibinfo{author}{\bibfnamefont{S.}~\bibnamefont{Tsujikawa}},
  \bibinfo{journal}{Int. J. Mod. Phys. D} \textbf{\bibinfo{volume}{15}},
  \bibinfo{pages}{1753} (\bibinfo{year}{2006}), \eprint{hep-th/0603057}.

\bibitem[{\citenamefont{Amendola and S.Tsujikawa}(2010)}]{amendola-book}
\bibinfo{author}{\bibfnamefont{L.}~\bibnamefont{Amendola}} \bibnamefont{and}
  \bibinfo{author}{\bibnamefont{S.Tsujikawa}} (\bibinfo{year}{2010}),
  \eprint{1912.07500}.

\bibitem[{\citenamefont{De~Felice and Tsujikawa}(2010)}]{DeFelice:2010aj}
\bibinfo{author}{\bibfnamefont{A.}~\bibnamefont{De~Felice}} \bibnamefont{and}
  \bibinfo{author}{\bibfnamefont{S.}~\bibnamefont{Tsujikawa}},
  \bibinfo{journal}{Living Rev. Rel.} \textbf{\bibinfo{volume}{13}},
  \bibinfo{pages}{3} (\bibinfo{year}{2010}), \eprint{1002.4928}.

\bibitem[{\citenamefont{Jaime et~al.}(2014{\natexlab{a}})\citenamefont{Jaime,
  Salgado, and Patino}}]{Jaime:2012nj}
\bibinfo{author}{\bibfnamefont{L.}~\bibnamefont{Jaime}},
  \bibinfo{author}{\bibfnamefont{M.}~\bibnamefont{Salgado}}, \bibnamefont{and}
  \bibinfo{author}{\bibfnamefont{L.}~\bibnamefont{Patino}},
  \bibinfo{journal}{Springer Proc. Phys.} \textbf{\bibinfo{volume}{157}},
  \bibinfo{pages}{363} (\bibinfo{year}{2014}{\natexlab{a}}),
  \eprint{1211.0015}.

\bibitem[{\citenamefont{Jaime et~al.}(2018)\citenamefont{Jaime, Jaber, and
  Escamilla-Rivera}}]{Jaime:2018ftn}
\bibinfo{author}{\bibfnamefont{L.~G.} \bibnamefont{Jaime}},
  \bibinfo{author}{\bibfnamefont{M.}~\bibnamefont{Jaber}}, \bibnamefont{and}
  \bibinfo{author}{\bibfnamefont{C.}~\bibnamefont{Escamilla-Rivera}},
  \bibinfo{journal}{Phys. Rev.} \textbf{\bibinfo{volume}{D98}},
  \bibinfo{pages}{083530} (\bibinfo{year}{2018}), \eprint{1804.04284}.

\bibitem[{\citenamefont{Jaime et~al.}(2014{\natexlab{b}})\citenamefont{Jaime,
  Patiño, and Salgado}}]{Jaime:2013zwa}
\bibinfo{author}{\bibfnamefont{L.~G.} \bibnamefont{Jaime}},
  \bibinfo{author}{\bibfnamefont{L.}~\bibnamefont{Patiño}}, \bibnamefont{and}
  \bibinfo{author}{\bibfnamefont{M.}~\bibnamefont{Salgado}},
  \bibinfo{journal}{Phys. Rev.} \textbf{\bibinfo{volume}{D89}},
  \bibinfo{pages}{084010} (\bibinfo{year}{2014}{\natexlab{b}}),
  \eprint{1312.5428}.

\bibitem[{\citenamefont{Jaime et~al.}(2011)\citenamefont{Jaime, Patino, and
  Salgado}}]{Jaime:2010kn}
\bibinfo{author}{\bibfnamefont{L.~G.} \bibnamefont{Jaime}},
  \bibinfo{author}{\bibfnamefont{L.}~\bibnamefont{Patino}}, \bibnamefont{and}
  \bibinfo{author}{\bibfnamefont{M.}~\bibnamefont{Salgado}},
  \bibinfo{journal}{Phys. Rev.} \textbf{\bibinfo{volume}{D83}},
  \bibinfo{pages}{024039} (\bibinfo{year}{2011}), \eprint{1006.5747}.

\bibitem[{\citenamefont{Starobinsky}(2007)}]{Starobinsky:2007}
\bibinfo{author}{\bibfnamefont{A.~A.} \bibnamefont{Starobinsky}},
  \bibinfo{journal}{JETP Lett.} \textbf{\bibinfo{volume}{86}},
  \bibinfo{pages}{1} (\bibinfo{year}{2007}).

\bibitem[{\citenamefont{Hu and Sawicki}(2007{\natexlab{a}})}]{Hu:2007}
\bibinfo{author}{\bibfnamefont{W.}~\bibnamefont{Hu}} \bibnamefont{and}
  \bibinfo{author}{\bibfnamefont{I.}~\bibnamefont{Sawicki}},
  \bibinfo{journal}{Phys. Rev. D} \textbf{\bibinfo{volume}{76}},
  \bibinfo{pages}{064004} (\bibinfo{year}{2007}{\natexlab{a}}).

\bibitem[{\citenamefont{Linder}(2009{\natexlab{a}})}]{Linder:2009}
\bibinfo{author}{\bibfnamefont{E.~V.} \bibnamefont{Linder}},
  \bibinfo{journal}{Phys. Rev. D} \textbf{\bibinfo{volume}{80}},
  \bibinfo{pages}{123528} (\bibinfo{year}{2009}{\natexlab{a}}),
  \eprint{0905.2962}.

\bibitem[{\citenamefont{Jaime and Salgado}(2018)}]{Jaime:2017gtr}
\bibinfo{author}{\bibfnamefont{L.~G.} \bibnamefont{Jaime}} \bibnamefont{and}
  \bibinfo{author}{\bibfnamefont{M.}~\bibnamefont{Salgado}},
  \bibinfo{journal}{Phys. Rev. D} \textbf{\bibinfo{volume}{98}},
  \bibinfo{pages}{084045} (\bibinfo{year}{2018}), \eprint{1711.08026}.

\bibitem[{\citenamefont{Odintsov et~al.}(2017)\citenamefont{Odintsov,
  Sáez-Chillón~Gómez, and Sharov}}]{Odintsov:2017qif}
\bibinfo{author}{\bibfnamefont{S.~D.} \bibnamefont{Odintsov}},
  \bibinfo{author}{\bibfnamefont{D.}~\bibnamefont{Sáez-Chillón~Gómez}},
  \bibnamefont{and} \bibinfo{author}{\bibfnamefont{G.~S.}
  \bibnamefont{Sharov}}, \bibinfo{journal}{Eur. Phys. J. C}
  \textbf{\bibinfo{volume}{77}}, \bibinfo{pages}{862} (\bibinfo{year}{2017}),
  \eprint{1709.06800}.

\bibitem[{\citenamefont{de~la Torre et~al.}(2017)}]{delaTorre:2016rxm}
\bibinfo{author}{\bibfnamefont{S.}~\bibnamefont{de~la Torre}}
  \bibnamefont{et~al.}, \bibinfo{journal}{Astron. Astrophys.}
  \textbf{\bibinfo{volume}{608}}, \bibinfo{pages}{A44} (\bibinfo{year}{2017}),
  \eprint{1612.05647}.

\bibitem[{\citenamefont{Schlegel et~al.}(2019)}]{Schlegel:2019eqc}
\bibinfo{author}{\bibfnamefont{D.~J.} \bibnamefont{Schlegel}}
  \bibnamefont{et~al.} (\bibinfo{year}{2019}), \eprint{1907.11171}.

\bibitem[{\citenamefont{Bowman et~al.}(2018)\citenamefont{Bowman, Rogers,
  Monsalve, Mozdzen, and Mahesh}}]{Bowman:2018yin}
\bibinfo{author}{\bibfnamefont{J.~D.} \bibnamefont{Bowman}},
  \bibinfo{author}{\bibfnamefont{A.~E.~E.} \bibnamefont{Rogers}},
  \bibinfo{author}{\bibfnamefont{R.~A.} \bibnamefont{Monsalve}},
  \bibinfo{author}{\bibfnamefont{T.~J.} \bibnamefont{Mozdzen}},
  \bibnamefont{and} \bibinfo{author}{\bibfnamefont{N.}~\bibnamefont{Mahesh}},
  \bibinfo{journal}{Nature} \textbf{\bibinfo{volume}{555}}, \bibinfo{pages}{67}
  (\bibinfo{year}{2018}), \eprint{1810.05912}.

\bibitem[{\citenamefont{Abbott et~al.}(2020)}]{LIGOScientific:2020stg}
\bibinfo{author}{\bibfnamefont{R.}~\bibnamefont{Abbott}} \bibnamefont{et~al.}
  (\bibinfo{collaboration}{LIGO Scientific, Virgo}) (\bibinfo{year}{2020}),
  \eprint{2004.08342}.

\bibitem[{\citenamefont{Corman et~al.}(2020)\citenamefont{Corman,
  Escamilla-Rivera, and Hendry}}]{Corman:2020pyr}
\bibinfo{author}{\bibfnamefont{M.}~\bibnamefont{Corman}},
  \bibinfo{author}{\bibfnamefont{C.}~\bibnamefont{Escamilla-Rivera}},
  \bibnamefont{and} \bibinfo{author}{\bibfnamefont{M.}~\bibnamefont{Hendry}}
  (\bibinfo{year}{2020}), \eprint{2004.04009}.

\bibitem[{\citenamefont{Capozziello and Tsujikawa}(2008)}]{Capozziello:2007eu}
\bibinfo{author}{\bibfnamefont{S.}~\bibnamefont{Capozziello}} \bibnamefont{and}
  \bibinfo{author}{\bibfnamefont{S.}~\bibnamefont{Tsujikawa}},
  \bibinfo{journal}{Phys. Rev. D} \textbf{\bibinfo{volume}{77}},
  \bibinfo{pages}{107501} (\bibinfo{year}{2008}), \eprint{0712.2268}.

\bibitem[{\citenamefont{Hu and Sawicki}(2007{\natexlab{b}})}]{Hu:2007nk}
\bibinfo{author}{\bibfnamefont{W.}~\bibnamefont{Hu}} \bibnamefont{and}
  \bibinfo{author}{\bibfnamefont{I.}~\bibnamefont{Sawicki}},
  \bibinfo{journal}{Phys. Rev. D} \textbf{\bibinfo{volume}{76}},
  \bibinfo{pages}{064004} (\bibinfo{year}{2007}{\natexlab{b}}),
  \eprint{0705.1158}.

\bibitem[{\citenamefont{Linder}(2009{\natexlab{b}})}]{Linder:2009jz}
\bibinfo{author}{\bibfnamefont{E.~V.} \bibnamefont{Linder}},
  \bibinfo{journal}{Phys. Rev. D} \textbf{\bibinfo{volume}{80}},
  \bibinfo{pages}{123528} (\bibinfo{year}{2009}{\natexlab{b}}),
  \eprint{0905.2962}.

\bibitem[{\citenamefont{Scolnic et~al.}(2018)}]{Scolnic:2017caz}
\bibinfo{author}{\bibfnamefont{D.~M.} \bibnamefont{Scolnic}}
  \bibnamefont{et~al.}, \bibinfo{journal}{Astrophys. J.}
  \textbf{\bibinfo{volume}{859}}, \bibinfo{pages}{101} (\bibinfo{year}{2018}),
  \eprint{1710.00845}.

\bibitem[{\citenamefont{Giostri et~al.}(2012)\citenamefont{Giostri, dos Santos,
  Waga, Reis, Calv{\~{a}}o, and Lago}}]{Giostri:2012}
\bibinfo{author}{\bibfnamefont{R.}~\bibnamefont{Giostri}},
  \bibinfo{author}{\bibfnamefont{M.~V.} \bibnamefont{dos Santos}},
  \bibinfo{author}{\bibfnamefont{I.}~\bibnamefont{Waga}},
  \bibinfo{author}{\bibfnamefont{R.}~\bibnamefont{Reis}},
  \bibinfo{author}{\bibfnamefont{M.}~\bibnamefont{Calv{\~{a}}o}},
  \bibnamefont{and} \bibinfo{author}{\bibfnamefont{B.~L.} \bibnamefont{Lago}},
  \bibinfo{journal}{Journal of Cosmology and Astroparticle Physics}
  \textbf{\bibinfo{volume}{2012}}, \bibinfo{pages}{027} (\bibinfo{year}{2012}),
  \urlprefix\url{https://doi.org/10.1088/1475-7516/2012/03/027}.

\bibitem[{\citenamefont{Percival et~al.}(2010)\citenamefont{Percival, Reid,
  Eisenstein, Bahcall, Budavari, Frieman, Fukugita, Gunn, Ivezic, Knapp
  et~al.}}]{Percival:2010}
\bibinfo{author}{\bibfnamefont{W.~J.} \bibnamefont{Percival}},
  \bibinfo{author}{\bibfnamefont{B.~A.} \bibnamefont{Reid}},
  \bibinfo{author}{\bibfnamefont{D.~J.} \bibnamefont{Eisenstein}},
  \bibinfo{author}{\bibfnamefont{N.~A.} \bibnamefont{Bahcall}},
  \bibinfo{author}{\bibfnamefont{T.}~\bibnamefont{Budavari}},
  \bibinfo{author}{\bibfnamefont{J.~A.} \bibnamefont{Frieman}},
  \bibinfo{author}{\bibfnamefont{M.}~\bibnamefont{Fukugita}},
  \bibinfo{author}{\bibfnamefont{J.~E.} \bibnamefont{Gunn}},
  \bibinfo{author}{\bibfnamefont{Z.}~\bibnamefont{Ivezic}},
  \bibinfo{author}{\bibfnamefont{G.~R.} \bibnamefont{Knapp}},
  \bibnamefont{et~al.}, \bibinfo{journal}{Monthly Notices of the Royal
  Astronomical Society} \textbf{\bibinfo{volume}{401}}, \bibinfo{pages}{2148}
  (\bibinfo{year}{2010}), ISSN \bibinfo{issn}{0035-8711},
  \eprint{https://academic.oup.com/mnras/article-pdf/401/4/2148/3901461/mnras0401-2148.pdf},
  \urlprefix\url{https://doi.org/10.1111/j.1365-2966.2009.15812.x}.

\bibitem[{\citenamefont{Blake et~al.}(2011)\citenamefont{Blake, Kazin, Beutler,
  Davis, Parkinson, Brough, Colless, Contreras, Couch, Croom
  et~al.}}]{Blake:2011}
\bibinfo{author}{\bibfnamefont{C.}~\bibnamefont{Blake}},
  \bibinfo{author}{\bibfnamefont{E.~A.} \bibnamefont{Kazin}},
  \bibinfo{author}{\bibfnamefont{F.}~\bibnamefont{Beutler}},
  \bibinfo{author}{\bibfnamefont{T.~M.} \bibnamefont{Davis}},
  \bibinfo{author}{\bibfnamefont{D.}~\bibnamefont{Parkinson}},
  \bibinfo{author}{\bibfnamefont{S.}~\bibnamefont{Brough}},
  \bibinfo{author}{\bibfnamefont{M.}~\bibnamefont{Colless}},
  \bibinfo{author}{\bibfnamefont{C.}~\bibnamefont{Contreras}},
  \bibinfo{author}{\bibfnamefont{W.}~\bibnamefont{Couch}},
  \bibinfo{author}{\bibfnamefont{S.}~\bibnamefont{Croom}},
  \bibnamefont{et~al.}, \bibinfo{journal}{Monthly Notices of the Royal
  Astronomical Society} \textbf{\bibinfo{volume}{418}}, \bibinfo{pages}{1707}
  (\bibinfo{year}{2011}), ISSN \bibinfo{issn}{0035-8711},
  \eprint{https://academic.oup.com/mnras/article-pdf/418/3/1707/18440857/mnras0418-1707.pdf},
  \urlprefix\url{https://doi.org/10.1111/j.1365-2966.2011.19592.x}.

\bibitem[{\citenamefont{{Beutler} et~al.}(2011)\citenamefont{{Beutler},
  {Blake}, {Colless}, {Jones}, {Staveley-Smith}, {Campbell}, {Parker},
  {Saunders}, and {Watson}}}]{Beutler:2011hx}
\bibinfo{author}{\bibfnamefont{F.}~\bibnamefont{{Beutler}}},
  \bibinfo{author}{\bibfnamefont{C.}~\bibnamefont{{Blake}}},
  \bibinfo{author}{\bibfnamefont{M.}~\bibnamefont{{Colless}}},
  \bibinfo{author}{\bibfnamefont{D.~H.} \bibnamefont{{Jones}}},
  \bibinfo{author}{\bibfnamefont{L.}~\bibnamefont{{Staveley-Smith}}},
  \bibinfo{author}{\bibfnamefont{L.}~\bibnamefont{{Campbell}}},
  \bibinfo{author}{\bibfnamefont{Q.}~\bibnamefont{{Parker}}},
  \bibinfo{author}{\bibfnamefont{W.}~\bibnamefont{{Saunders}}},
  \bibnamefont{and} \bibinfo{author}{\bibfnamefont{F.}~\bibnamefont{{Watson}}},
  \bibinfo{journal}{mnras} \textbf{\bibinfo{volume}{416}},
  \bibinfo{pages}{3017} (\bibinfo{year}{2011}), \eprint{1106.3366}.

\bibitem[{\citenamefont{Jimenez and Loeb}(2002)}]{Jimenez:2001gg}
\bibinfo{author}{\bibfnamefont{R.}~\bibnamefont{Jimenez}} \bibnamefont{and}
  \bibinfo{author}{\bibfnamefont{A.}~\bibnamefont{Loeb}},
  \bibinfo{journal}{Astrophys. J.} \textbf{\bibinfo{volume}{573}},
  \bibinfo{pages}{37} (\bibinfo{year}{2002}), \eprint{astro-ph/0106145}.

\bibitem[{\citenamefont{Amante et~al.}(2020)\citenamefont{Amante, Maga\~na,
  Motta, Garc\'\i{}a-Aspeitia, and Verdugo}}]{Amante:2019xao}
\bibinfo{author}{\bibfnamefont{M.~H.} \bibnamefont{Amante}},
  \bibinfo{author}{\bibfnamefont{J.}~\bibnamefont{Maga\~na}},
  \bibinfo{author}{\bibfnamefont{V.}~\bibnamefont{Motta}},
  \bibinfo{author}{\bibfnamefont{M.~A.} \bibnamefont{Garc\'\i{}a-Aspeitia}},
  \bibnamefont{and} \bibinfo{author}{\bibfnamefont{T.}~\bibnamefont{Verdugo}},
  \bibinfo{journal}{Mon. Not. Roy. Astron. Soc.}
  \textbf{\bibinfo{volume}{498}}, \bibinfo{pages}{6013} (\bibinfo{year}{2020}),
  \eprint{1906.04107}.

\bibitem[{\citenamefont{Biesiada}(2006)}]{Biesiada:2006zf}
\bibinfo{author}{\bibfnamefont{M.}~\bibnamefont{Biesiada}},
  \bibinfo{journal}{Phys. Rev.} \textbf{\bibinfo{volume}{D73}},
  \bibinfo{pages}{023006} (\bibinfo{year}{2006}).

\bibitem[{\citenamefont{{Biesiada} et~al.}(2010)\citenamefont{{Biesiada},
  {Pi{\'o}rkowska}, and {Malec}}}]{Biesiada:2010}
\bibinfo{author}{\bibfnamefont{M.}~\bibnamefont{{Biesiada}}},
  \bibinfo{author}{\bibfnamefont{A.}~\bibnamefont{{Pi{\'o}rkowska}}},
  \bibnamefont{and} \bibinfo{author}{\bibfnamefont{B.}~\bibnamefont{{Malec}}},
  \bibinfo{journal}{mnras} \textbf{\bibinfo{volume}{406}},
  \bibinfo{pages}{1055} (\bibinfo{year}{2010}), \eprint{1105.0946}.

\bibitem[{\citenamefont{{Cao} et~al.}(2012)\citenamefont{{Cao}, {Pan},
  {Biesiada}, {Godlowski}, and {Zhu}}}]{Cao:2012}
\bibinfo{author}{\bibfnamefont{S.}~\bibnamefont{{Cao}}},
  \bibinfo{author}{\bibfnamefont{Y.}~\bibnamefont{{Pan}}},
  \bibinfo{author}{\bibfnamefont{M.}~\bibnamefont{{Biesiada}}},
  \bibinfo{author}{\bibfnamefont{W.}~\bibnamefont{{Godlowski}}},
  \bibnamefont{and} \bibinfo{author}{\bibfnamefont{Z.-H.} \bibnamefont{{Zhu}}},
  \bibinfo{journal}{jcap} \textbf{\bibinfo{volume}{3}}, \bibinfo{eid}{016}
  (\bibinfo{year}{2012}), \eprint{1105.6226}.

\bibitem[{\citenamefont{Cao et~al.}(2015)\citenamefont{Cao, Biesiada, Gavazzi,
  Piórkowska, and Zhu}}]{Cao:2015qja}
\bibinfo{author}{\bibfnamefont{S.}~\bibnamefont{Cao}},
  \bibinfo{author}{\bibfnamefont{M.}~\bibnamefont{Biesiada}},
  \bibinfo{author}{\bibfnamefont{R.}~\bibnamefont{Gavazzi}},
  \bibinfo{author}{\bibfnamefont{A.}~\bibnamefont{Piórkowska}},
  \bibnamefont{and} \bibinfo{author}{\bibfnamefont{Z.-H.} \bibnamefont{Zhu}},
  \bibinfo{journal}{Astrophys. J.} \textbf{\bibinfo{volume}{806}},
  \bibinfo{pages}{185} (\bibinfo{year}{2015}), \eprint{1509.07649}.

\bibitem[{\citenamefont{{Planck Collaboration}
  et~al.}(2020{\natexlab{a}})\citenamefont{{Planck Collaboration}, {Aghanim,
  N.}, {Akrami, Y.}, {Arroja, F.}, {Ashdown, M.}, {Aumont, J.}, {Baccigalupi,
  C.}, {Ballardini, M.}, {Banday, A. J.}, {Barreiro, R. B.}
  et~al.}}]{Planck_overview:2018}
\bibinfo{author}{\bibnamefont{{Planck Collaboration}}},
  \bibinfo{author}{\bibnamefont{{Aghanim, N.}}},
  \bibinfo{author}{\bibnamefont{{Akrami, Y.}}},
  \bibinfo{author}{\bibnamefont{{Arroja, F.}}},
  \bibinfo{author}{\bibnamefont{{Ashdown, M.}}},
  \bibinfo{author}{\bibnamefont{{Aumont, J.}}},
  \bibinfo{author}{\bibnamefont{{Baccigalupi, C.}}},
  \bibinfo{author}{\bibnamefont{{Ballardini, M.}}},
  \bibinfo{author}{\bibnamefont{{Banday, A. J.}}},
  \bibinfo{author}{\bibnamefont{{Barreiro, R. B.}}}, \bibnamefont{et~al.},
  \bibinfo{journal}{A\&A} \textbf{\bibinfo{volume}{641}}, \bibinfo{pages}{A1}
  (\bibinfo{year}{2020}{\natexlab{a}}),
  \urlprefix\url{https://doi.org/10.1051/0004-6361/201833880}.

\bibitem[{\citenamefont{{Planck Collaboration}
  et~al.}(2020{\natexlab{b}})\citenamefont{{Planck Collaboration}, {Aghanim,
  N.}, {Akrami, Y.}, {Ashdown, M.}, {Aumont, J.}, {Baccigalupi, C.},
  {Ballardini, M.}, {Banday, A. J.}, {Barreiro, R. B.}, {Bartolo, N.}
  et~al.}}]{Planck_like:2018}
\bibinfo{author}{\bibnamefont{{Planck Collaboration}}},
  \bibinfo{author}{\bibnamefont{{Aghanim, N.}}},
  \bibinfo{author}{\bibnamefont{{Akrami, Y.}}},
  \bibinfo{author}{\bibnamefont{{Ashdown, M.}}},
  \bibinfo{author}{\bibnamefont{{Aumont, J.}}},
  \bibinfo{author}{\bibnamefont{{Baccigalupi, C.}}},
  \bibinfo{author}{\bibnamefont{{Ballardini, M.}}},
  \bibinfo{author}{\bibnamefont{{Banday, A. J.}}},
  \bibinfo{author}{\bibnamefont{{Barreiro, R. B.}}},
  \bibinfo{author}{\bibnamefont{{Bartolo, N.}}}, \bibnamefont{et~al.},
  \bibinfo{journal}{A\&A} \textbf{\bibinfo{volume}{641}}, \bibinfo{pages}{A5}
  (\bibinfo{year}{2020}{\natexlab{b}}),
  \urlprefix\url{https://doi.org/10.1051/0004-6361/201936386}.

\bibitem[{\citenamefont{Escamilla-Rivera
  et~al.}(2016)\citenamefont{Escamilla-Rivera, Casarini, Fabris, and
  Alcaniz}}]{Escamilla-Rivera:2016aca}
\bibinfo{author}{\bibfnamefont{C.}~\bibnamefont{Escamilla-Rivera}},
  \bibinfo{author}{\bibfnamefont{L.}~\bibnamefont{Casarini}},
  \bibinfo{author}{\bibfnamefont{J.~C.} \bibnamefont{Fabris}},
  \bibnamefont{and} \bibinfo{author}{\bibfnamefont{J.~S.}
  \bibnamefont{Alcaniz}}, \bibinfo{journal}{JCAP}
  \textbf{\bibinfo{volume}{11}}, \bibinfo{pages}{010} (\bibinfo{year}{2016}),
  \eprint{1605.01475}.

\bibitem[{\citenamefont{Kazantzidis and
  Perivolaropoulos}(2018)}]{Kazantzidis:2018rnb}
\bibinfo{author}{\bibfnamefont{L.}~\bibnamefont{Kazantzidis}} \bibnamefont{and}
  \bibinfo{author}{\bibfnamefont{L.}~\bibnamefont{Perivolaropoulos}},
  \bibinfo{journal}{Phys. Rev. D} \textbf{\bibinfo{volume}{97}},
  \bibinfo{pages}{103503} (\bibinfo{year}{2018}), \eprint{1803.01337}.

\bibitem[{\citenamefont{Verde et~al.}(2019)\citenamefont{Verde, Treu, and
  Riess}}]{Verde:2019ivm}
\bibinfo{author}{\bibfnamefont{L.}~\bibnamefont{Verde}},
  \bibinfo{author}{\bibfnamefont{T.}~\bibnamefont{Treu}}, \bibnamefont{and}
  \bibinfo{author}{\bibfnamefont{A.}~\bibnamefont{Riess}}
  (\bibinfo{year}{2019}), \eprint{1907.10625}.

\bibitem[{\citenamefont{Battye et~al.}(2015)\citenamefont{Battye, Charnock, and
  Moss}}]{Battye:2014qga}
\bibinfo{author}{\bibfnamefont{R.~A.} \bibnamefont{Battye}},
  \bibinfo{author}{\bibfnamefont{T.}~\bibnamefont{Charnock}}, \bibnamefont{and}
  \bibinfo{author}{\bibfnamefont{A.}~\bibnamefont{Moss}},
  \bibinfo{journal}{Phys. Rev.} \textbf{\bibinfo{volume}{D91}},
  \bibinfo{pages}{103508} (\bibinfo{year}{2015}), \eprint{1409.2769}.

\bibitem[{\citenamefont{Mccarthy et~al.}(2018)\citenamefont{Mccarthy, Bird,
  Schaye, Harnois-Deraps, Font, and Van~Waerbeke}}]{McCarthy:2017csu}
\bibinfo{author}{\bibfnamefont{I.~G.} \bibnamefont{Mccarthy}},
  \bibinfo{author}{\bibfnamefont{S.}~\bibnamefont{Bird}},
  \bibinfo{author}{\bibfnamefont{J.}~\bibnamefont{Schaye}},
  \bibinfo{author}{\bibfnamefont{J.}~\bibnamefont{Harnois-Deraps}},
  \bibinfo{author}{\bibfnamefont{A.~S.} \bibnamefont{Font}}, \bibnamefont{and}
  \bibinfo{author}{\bibfnamefont{L.}~\bibnamefont{Van~Waerbeke}},
  \bibinfo{journal}{Mon. Not. Roy. Astron. Soc.}
  \textbf{\bibinfo{volume}{476}}, \bibinfo{pages}{2999} (\bibinfo{year}{2018}),
  \eprint{1712.02411}.

\bibitem[{\citenamefont{Di~Valentino et~al.}(2020)}]{DiValentino:2020vvd}
\bibinfo{author}{\bibfnamefont{E.}~\bibnamefont{Di~Valentino}}
  \bibnamefont{et~al.} (\bibinfo{year}{2020}), \eprint{2008.11285}.

\bibitem[{\citenamefont{{Planck Collaboration}
  et~al.}(2014)\citenamefont{{Planck Collaboration}, {Ade}, {Aghanim},
  {Armitage-Caplan}, {Arnaud}, {Ashdown}, {Atrio-Barandela}, {Aumont},
  {Baccigalupi}, {Banday} et~al.}}]{2014A&A...571A..20P}
\bibinfo{author}{\bibnamefont{{Planck Collaboration}}},
  \bibinfo{author}{\bibfnamefont{P.~A.~R.} \bibnamefont{{Ade}}},
  \bibinfo{author}{\bibfnamefont{N.}~\bibnamefont{{Aghanim}}},
  \bibinfo{author}{\bibfnamefont{C.}~\bibnamefont{{Armitage-Caplan}}},
  \bibinfo{author}{\bibfnamefont{M.}~\bibnamefont{{Arnaud}}},
  \bibinfo{author}{\bibfnamefont{M.}~\bibnamefont{{Ashdown}}},
  \bibinfo{author}{\bibfnamefont{F.}~\bibnamefont{{Atrio-Barandela}}},
  \bibinfo{author}{\bibfnamefont{J.}~\bibnamefont{{Aumont}}},
  \bibinfo{author}{\bibfnamefont{C.}~\bibnamefont{{Baccigalupi}}},
  \bibinfo{author}{\bibfnamefont{A.~J.} \bibnamefont{{Banday}}},
  \bibnamefont{et~al.}, \bibinfo{journal}{APP} \textbf{\bibinfo{volume}{571}},
  \bibinfo{eid}{A20} (\bibinfo{year}{2014}), \eprint{1303.5080}.

\bibitem[{\citenamefont{{DES Collaboration} et~al.}(2017)\citenamefont{{DES
  Collaboration}, {Abbott}, {Abdalla}, {Alarcon}, {Aleksi{\'c}}, {Allam},
  {Allen}, {Amara}, {Annis}, {Asorey} et~al.}}]{2017arXiv170801530D}
\bibinfo{author}{\bibnamefont{{DES Collaboration}}},
  \bibinfo{author}{\bibfnamefont{T.~M.~C.} \bibnamefont{{Abbott}}},
  \bibinfo{author}{\bibfnamefont{F.~B.} \bibnamefont{{Abdalla}}},
  \bibinfo{author}{\bibfnamefont{A.}~\bibnamefont{{Alarcon}}},
  \bibinfo{author}{\bibfnamefont{J.}~\bibnamefont{{Aleksi{\'c}}}},
  \bibinfo{author}{\bibfnamefont{S.}~\bibnamefont{{Allam}}},
  \bibinfo{author}{\bibfnamefont{S.}~\bibnamefont{{Allen}}},
  \bibinfo{author}{\bibfnamefont{A.}~\bibnamefont{{Amara}}},
  \bibinfo{author}{\bibfnamefont{J.}~\bibnamefont{{Annis}}},
  \bibinfo{author}{\bibfnamefont{J.}~\bibnamefont{{Asorey}}},
  \bibnamefont{et~al.}, \bibinfo{journal}{ArXiv e-prints}
  \bibinfo{eid}{arXiv:1708.01530} (\bibinfo{year}{2017}), \eprint{1708.01530}.

\bibitem[{\citenamefont{Hildebrandt et~al.}(2017)}]{Hildebrandt:2016iqg}
\bibinfo{author}{\bibfnamefont{H.}~\bibnamefont{Hildebrandt}}
  \bibnamefont{et~al.}, \bibinfo{journal}{Mon. Not. Roy. Astron. Soc.}
  \textbf{\bibinfo{volume}{465}}, \bibinfo{pages}{1454} (\bibinfo{year}{2017}),
  \eprint{1606.05338}.

\bibitem[{\citenamefont{Joudaki et~al.}(2017)}]{Joudaki:2016mvz}
\bibinfo{author}{\bibfnamefont{S.}~\bibnamefont{Joudaki}} \bibnamefont{et~al.},
  \bibinfo{journal}{Mon. Not. Roy. Astron. Soc.}
  \textbf{\bibinfo{volume}{465}}, \bibinfo{pages}{2033} (\bibinfo{year}{2017}),
  \eprint{1601.05786}.

\bibitem[{\citenamefont{Zhao et~al.}(2018)\citenamefont{Zhao, Zhang, and
  Zhang}}]{Zhao:2017jma}
\bibinfo{author}{\bibfnamefont{M.-M.} \bibnamefont{Zhao}},
  \bibinfo{author}{\bibfnamefont{J.-F.} \bibnamefont{Zhang}}, \bibnamefont{and}
  \bibinfo{author}{\bibfnamefont{X.}~\bibnamefont{Zhang}},
  \bibinfo{journal}{Phys. Lett. B} \textbf{\bibinfo{volume}{779}},
  \bibinfo{pages}{473} (\bibinfo{year}{2018}), \eprint{1710.02391}.

\bibitem[{\citenamefont{Capozziello et~al.}(2008)\citenamefont{Capozziello,
  Cardone, and Salzano}}]{Capozziello:2008qc}
\bibinfo{author}{\bibfnamefont{S.}~\bibnamefont{Capozziello}},
  \bibinfo{author}{\bibfnamefont{V.~F.} \bibnamefont{Cardone}},
  \bibnamefont{and} \bibinfo{author}{\bibfnamefont{V.}~\bibnamefont{Salzano}},
  \bibinfo{journal}{Phys. Rev. D} \textbf{\bibinfo{volume}{78}},
  \bibinfo{pages}{063504} (\bibinfo{year}{2008}), \eprint{0802.1583}.

\bibitem[{\citenamefont{Capozziello et~al.}(2019)\citenamefont{Capozziello,
  D'Agostino, and Luongo}}]{Capozziello:2019cav}
\bibinfo{author}{\bibfnamefont{S.}~\bibnamefont{Capozziello}},
  \bibinfo{author}{\bibfnamefont{R.}~\bibnamefont{D'Agostino}},
  \bibnamefont{and} \bibinfo{author}{\bibfnamefont{O.}~\bibnamefont{Luongo}},
  \bibinfo{journal}{Int. J. Mod. Phys. D} \textbf{\bibinfo{volume}{28}},
  \bibinfo{pages}{1930016} (\bibinfo{year}{2019}), \eprint{1904.01427}.

\bibitem[{\citenamefont{Capozziello et~al.}(2015)\citenamefont{Capozziello,
  Luongo, and Saridakis}}]{Capozziello:2015rda}
\bibinfo{author}{\bibfnamefont{S.}~\bibnamefont{Capozziello}},
  \bibinfo{author}{\bibfnamefont{O.}~\bibnamefont{Luongo}}, \bibnamefont{and}
  \bibinfo{author}{\bibfnamefont{E.~N.} \bibnamefont{Saridakis}},
  \bibinfo{journal}{Phys. Rev. D} \textbf{\bibinfo{volume}{91}},
  \bibinfo{pages}{124037} (\bibinfo{year}{2015}), \eprint{1503.02832}.

\end{thebibliography}

\end{document}